# Three-dimensional flat band in ultra-thin Kagome metal $Mn_3Sn$ film

Mengting Zhao[1,⊥,*], James Blyth[1,⊥], Grace L. Causer[1], Hongrun Zhang[2], Tianye Yu[3], Jiayu Liu[4], Wenchuan Jin[4], Mohammad T. H. Bhuiyan[1], Zheng-Tai Liu[4], Mao Ye[4], Yi Du[2], Zhiping Yin[5], Michael S. Fuhrer[1], Anton Tadich[6], Mark T. Edmonds[1,*]

[1] School of Physics and Astronomy, Monash University, Clayton, VIC 3800 Australia

[2] School of Physics, Beihang University, Haidian District, Beijing 100191, China

[3] Shenyang National Laboratory for Materials Science, Institute of Metal Research, Chinese Academy of Sciences, Shenyang 110016, China

[4] Shanghai Synchrotron Radiation Facility, Shanghai Advanced Research Institute, Chinese Academy of Sciences, Shanghai 201210, China

[5] School of Physics & Astronomy and Center for Advanced Quantum Studies, Beijing Normal University, Beijing 100875, China

[6] Australian Synchrotron, Clayton, VIC 3168 Australia

[⊥] These authors contributed equally to this work

\* Corresponding author: mengting.zhao@monash.edu and mark.edmonds@monash.edu

**Abstract**

Flat bands with small energy dispersion can give rise to strongly correlated electronic and topological phases, especially when located at the Fermi level. Whilst flat bands have been experimentally realized in two-dimensional (2D) twisted van der Waals heterostructures, they are highly sensitive to twist angle, necessitating complex fabrication techniques. Geometrically frustrated kagome lattices have emerged as an attractive platform as they natively host flat bands that have been observed experimentally in quasi-2D bulk-crystal kagome metals. An outstanding experimental question is whether flat bands can be realized in atomically thin metals, with opportunities for stronger electron-electron interactions through tuning of the surrounding dielectric environment. Here we use angle-resolved photoelectron spectroscopy, scanning tunnelling microscopy and band structure calculations to show that ultra-thin films of the kagome metal $Mn_3Sn$ host a robust dispersionless flat band with a bandwidth of 50 meV.

Furthermore, we demonstrate chemical tuning of the flat band to near the Fermi level via manganese defect engineering. The realization of tunable kagome-derived flat bands in an ultra-thin kagome metal, represents a promising platform to study strongly correlated and topological phenomena, with applications in quantum computing, spintronics and low-energy electronics.

Quantum materials based upon the kagome crystalline lattice have recently emerged as an abundant landscape of quantum phases driven by the interplay of topology, magnetism and strong electron correlations[1–6]. The rich physics in the kagome system derives from its unique corner-sharing triangular geometry and hexagonal symmetry. As shown in the left panel of Fig.1a, in a two-dimensional (2D) kagome lattice, the eigenfunctions at neighbouring nodes host opposite phases which destructively interfere to prevent electron hopping to the third node, and thus electronic states are confined within hexagonal plaquettes leading to flat band states[7] as depicted in the right panel of Fig. 1a. The flat band is also accompanied by dispersive bands with Dirac/Weyl points and the van Hove singularities that arise from the hexagonal structure. The quantum destructive interference quenches the electron's velocity and thus localizes it in real space, resulting in an extremely large density of states (DOS) of 'heavy' electrons in a narrow energy scale. In a similar manner to Landau levels, the strong electron correlations in flat bands can induce high-temperature superconductivity[8], the fractional quantum Hall effect[9,10], unconventional ferromagnetism[11], and Mott insulator phases[12,13]. Interestingly, constructing a kagome lattice on each face of a polyhedron, such as in a pyrochlore network, can induce three-dimensional (3D) destructive interference of electrons, resulting in 3D flat band states at every out-of-plane ($k_z$) momentum position[14–16]. This could potentially give rise to new higher-order correlated phenomena including fractionalized topological states[17], but to date, experimental realization of 3D flat bands is limited [14,15], and whether they can be realized in a quasi-2D kagome material remains elusive.

Ultra-thin kagome metals such as T$_m$X$_n$ (T = Fe, Co, Mn; X = Ge, Sn; m:n = 3:2, 3:1, 1:1) and RMn$_6$Sn$_6$ (R presents rare earth elements) have emerged as one of the most promising systems to study the interplay of topology, magnetism and correlations in the kagome lattice. Bulk kagome metals host linearly dispersive Dirac/Weyl bands and robust flat bands[18–21]. Whereas it is predicted that reducing the dimensionality of a magnetic kagome metal to monolayer and few-layers yields a non-trivial Chern insulator with a chiral edge state that could support an intrinsic quantum anomalous Hall effect and lossless transport of electrons[22]. Further, the fractional quantum Hall state is also predicted to arise in monolayer and few-layer kagome systems due to the flat band feature[23]. However, whether flat band states persist in ultra-thin kagome metals remains elusive to experimental observation. Realising few-layer or ultra-thin kagome systems provides a platform to exploit these kagome-lattice derived strongly correlated topological and quantum phases[20], as well magnetic phases such as Stoner-type ferromagnetism[24,25], as the flat band can be tuned to the Fermi energy ($E_F$) via electrostatic gating, doping and substrate engineering. However, experiments on few-layer kagome metals are lacking at present, with only >10 nm thin films grown to date[25,26]. To unambiguously demonstrate the flat band in ultrathin (3 nm) $Mn_3Sn$, we employ two independent experimental techniques to probe its dispersion in both momentum and real space. First, we utilize angle-resolved photoelectron spectroscopy (ARPES) to comprehensively map the 3D electronic band structure and find a prominent flat band near $E_F$, which is dispersionless in $k_z$, indicating an overall 3D flat band character. Second, we use scanning tunnelling spectroscopy (STS), which measures the local density of states (LDOS) as a function of energy, to probe the flat band in real space. These experimental observations of a robust 3D flat band are in excellent agreement with density-functional theory (DFT) and dynamical mean-field theory (DMFT) band structure calculations.

$Mn_3Sn$ is a hexagonal material that forms in the space group of *$P6_3/mmc$*, with in-plane and out-of-plane lattice constants of 0.566 nm, and 0.453 nm, respectively[27]. As shown in Fig.1b, the Mn atoms form a kagome lattice in each layer, and the Sn atoms form a hexagonal sublattice where each Sn atom is situated in the centre of the kagome lattice for chemical stability. Ultra-thin $Mn_3Sn$ films were successfully grown using molecular beam epitaxy (MBE) down to 3 nm thickness on Si (111)7×7 (see Methods for growth details). Reflection high-energy electron diffraction (RHEED) in Fig. 1c, reveals sharp and uniform streaks indicating a high-quality epitaxial $Mn_3Sn$ film (the expected six-fold crystal structure is confirmed by low-energy electron diffraction (Fig. S1)). The $Mn_3Sn$ chemical stoichiometry was confirmed using synchrotron-based X-ray photoelectron spectroscopy (XPS) at $hv$ = 880 eV, as shown in Fig. 1d. No additional components were observed in either the Mn and Sn core levels of $Mn_3Sn$ or the Si 2*p* core level (see Supplementary information Fig. S1a) corresponding to the Si(111) substrate, verifying that $Mn_3Sn$ is free-standing on Si(111). Large area scanning tunnelling microscopy (STM) reveals 3–5 nm thick $Mn_3Sn$ islands, that are atomically flat and up to 30 nm in size, along with small areas of bare substrate as shown in Fig. S1b. Figure 1e shows STM topography taken across a step edge, revealing a step height of ~2.4 Å, which corresponds to $c/2$ of the $Mn_3Sn$ lattice as labelled in Fig.1b. Figure 1f shows an atomically resolved STM image of the $Mn_3Sn$ film surface, revealing a honeycomb lattice and the expected in-plane lattice constant of 5.7 Å, consistent with previous studies on bulk $Mn_3Sn$[28,29].

Figure 2 details the electronic band structure results on ultra-thin $Mn_3Sn$ using ARPES. Considering the 3D Brillouin Zone (BZ) of bulk $Mn_3Sn$ (Fig. 2a), we used photon energy dependent ARPES to vary the out-of-plane momentum $k_z$ to map the electronic structure and identify the bulk Γ and A symmetry points (see Section 2 of SI for discussion). Figure 2 shows ARPES data acquired at a temperature $T = 11$ K (i.e. below the Néel temperature $T_N = 30$ K, see Section 3 of SI) using $hv = 104$ eV where the final state cuts approximately through the Γ-M-K $k_z$ plane. As shown in Fig. 2b-c ARPES spectra taken along the Γ-M and Γ-K high-symmetry directions display a prominent nondispersive band near the Fermi level at –0.3 eV. A sharp inverted V-shaped band due to the bulk valence band of Si(111) is also present at –1.5 eV[30,31]. The flat band remains essentially dispersionless throughout most of the BZ, except for an increased intensity at Γ caused by an intersecting parabolic band (see theory calculation in Fig. 2e). Figure 2d shows constant energy contours ($k_x$-$k_y$ mapping at $hv = 104$ eV) taken above the flat band (–0.1 eV) which shows the characteristic hexagonal six-fold symmetry of the $Mn_3Sn$ kagome lattice (BZ is overlaid in red). At the flat band (–0.3 eV), the spectral weight is significantly higher, and is mostly uniform across the entire BZ, further confirming the flat band exists across the whole BZ in agreement with theoretical predictions[32]. The bandwidth of the flat band over the entire BZ does not exceed 150 meV, indicating that quantum interference effects significantly suppress the electron kinetic energy, thereby hindering the wave function from spreading across the lattice. The flat band state is also relatively independent of temperature and magnetic order, as shown in Fig. S5 with a flat band observed in ARPES at $T = 78$ K which is well above $T_N = 30$ K.

To understand the origin of the observed flat band, in Fig. 2e we use density functional theory combined with dynamical mean-field theory (DFT + DMFT) to calculate the band structure of non-collinear antiferromagnetic (NAFM) $Mn_3Sn$. The NAFM state in ultra-thin $Mn_3Sn$ was confirmed from temperature dependent magnetization measurements (see SI Section 3 for results and discussion) and is consistent with thick film and bulk $Mn_3Sn$[33–36]. Overall, the calculations are in good agreement with the experimental observation of the flat band (red arrows) which extends across most of the BZ. A second flat band appears at –0.8 eV (pink arrow) that is not observed experimentally, possibly due to its very low spectral weight and matrix element effects. At the same time strongly dispersing Weyl bands emerge close to the Fermi level around the K points with the Weyl points marked by white arrows. It should be noted that DFT calculations of the electronic structure in the nonmagnetic phase of $Mn_3Sn$[27] also predict a flat band state suggesting that the origin of the flat band is due to kagome lattice frustration, supporting our experimental observation of the flat band persisting well above $T_N$ (Fig. S5). Additionally, we demonstrate that the position of the flat band is tunable. Figure S6 plots a series of ultra-thin $Mn_3Sn$ films with increasing amounts of Mn vacancies to demonstrate that the flat band can be shifted ~300 meV towards the Fermi level by slight tuning of the growth process to induce Mn vacancies, which results in hole-doping. This is consistent with previous studies on bulk $Mn_3Sn$, extra Mn concentration induced the electron doping and the same Fermi level shifting phenomenon has been proved by ARPES results as well as DFT calculations[37]. Also, electron/hole doping phenomenon has also been observed in Fe/Cr doped bulk $Mn_3Sn$ samples[38], highlighting flexibility of Fermi level of $Mn_3Sn$ via modifying $3d$ electrons composition.

Photon-energy dependent ARPES was utilized to probe the dimensionality (2D or 3D) of the flat band[14,15]. In Fig. 3a, we show the progression of the flat band as a function of photon energy (i.e. changing out-of-plane momentum $k_z$) from 88 eV to 128 eV corresponding to sampling the bandstructure within the Γ–M–L–A plane in momentum space. The position of the flat band appears to be robust to changes in photon energy, only varying in intensity, whilst the inverted V-shaped band from the Si(111) is strongly dispersive, as expected. To explore the flat band dispersion in more detail, Fig. 3b plots Energy vs out-of-plane momentum $k_z$, where the pronounced flat band near the Fermi level displays negligible dispersion (<100 meV) along Γ–A (from $k_z = 4.48$ Å$^{-1}$ to $k_z = 5.02$ Å$^{-1}$). Even though our samples are thin, they contain 13-20 kagome layers, and would therefore be expected to support a 3D dispersion if one existed. Coherence across the thickness of the film would create a series of 2D quantum-well states, each dispersionless, however these states would each inherit discrete momenta and energies from the $k_z$ dispersion. The fact that we observe no discrete energy or momentum states thus rules out any dispersion along $k_z$.

To understand the origin of the 3D flat band, Fig. S7 shows DFT+DMFT band structure calculations along the L-A-Γ-K direction. Focussing on the region near the Fermi level (marked by the yellow rectangle), minimal band dispersion is observed along the A–Γ direction, which contrasts with the significant dispersion along the L–A direction. According to previous theoretical study in Mn$_3$Sn, the dispersionless band near the Fermi level along the Γ–A direction arises from the suppression of magnetic fluctuations and the associated many-body renormalizations (e.g., quasiparticle mass enhancement) induced by magnetic order in Mn$_3$Sn, revealed by comparing DFT and DFT+DMFT calculations[27]. Fig. 3c plots the out-of-plane ARPES results overlaid with the calculated band structure along Γ–A, highlighting the good agreement. For a comparison with the A-L direction, we can examine ARPES spectra taken at $hv$ = 128 eV in Fig. 3a, which represents a cut nearly along the A–L direction which reveals a minimally dispersive flat band. This is in direct contrast with the dispersive band structure predicted for bulk Mn$_3$Sn in Fig. S7. This suggests a new mechanism of 3D flat band formation—specifically, that the quasi-2D nature of our sample enables the realization of a true 3D flat band in ultra-thin Mn$_3$Sn. To date, only several 3D flat band states have been realized experimentally, for example, the pyrochlore materials such as CaNi$_2$[14], CuV$_2$S$_4$[15], as well as the bulk kagome metal CoSn[20,39]. The former results from their 3D geometrically frustrated lattices, and the latter from weak interlayer interactions between the kagome layers due to the isolation of each Co$_3$Sn kagome layer by a Sn$_2$ stanene layer which results in a quasi-2D structure. In Mn$_3$Sn, the kagome layers stack directly on top of each other (Fig. 1b), and thus the $k_z$-dispersionless flat band in our Mn$_3$Sn film likely results from its quasi-2D formation (only 3 nm-thick) that suppresses interlayer hopping and its strong electron correlations.

We now probe the electronic properties of ultra-thin Mn$_3$Sn at the nanoscale using differential conductance spectroscopy (d$I$/d$V$) measurements (Fig. 4) to understand the spatial dependence of the flat band. Figure 4a is an atomic resolution STM image with the expected hexagonal lattice arrangement of the region where d$I$/d$V$ mapping was conducted. Figure 4b, is an area-averaged d$I$/d$V$ curve and exhibits three characteristic peaks, a prominent peak located at –25 mV, accompanied by two additional weaker peaks at –200 mV and –400 mV, respectively. We further probed the spatial distribution of these electronic states with a 2D contour plot of d$I$/d$V$ spectra acquired across the whole region in Fig. 4c. The prominent peak at –25 mV exhibits negligible dispersion and significant intensity, indicative of a substantial local DOS consistent with the flat band characteristics previously observed via STS in other correlated systems[21,40]. The full width at half maximum (FWHM) of the flat band peak is ~50 mV, which is significantly narrower than the ~150 meV bandwidth extracted from ARPES in Fig. 2. The increased broadening in ARPES may be caused by averaging effects over multiple domains and thicknesses due to the beam spot size, or band broadening that is common in thin-films measured with ARPES. The bandwidth of the flat band observed here is significantly narrower than that in other kagome metals, such as CoSn (~200 meV)[19], FeSn (~100 meV)[18,21], and YMn$_6$Sn$_6$ (~150 meV)[40], as well as in twisted silicene (100 meV)[41], and is only comparable with twisted bilayer graphene (~ 50 meV)[42]. The d$I$/d$V$ spectra in Fig. 4b also exhibits a distinct minimum in the LDOS located at +50 mV due to the Weyl points in reasonable agreement with theory in Fig. 2e. In STS experiments, bias-dependent d$I$/d$V$ mapping can be used to further verify flat band states, as the electron distribution is strongly localized on the Mn sites corresponding to the kagome lattice only at the flat band energy. Also, STS mapping confirms that the flat band states are uniform across the sample, and not due to e.g. defect states. Figure 4d-f show d$I$/d$V$ maps obtained from the same region as Fig. 4a acquired at the characteristic peak locations identified in the Fig. 4b i.e. –400 mV, –200 mV and the flat band energy at –25 mV. To clearly recognise which atoms contribute to the LDOS at each specific energy, a simple Mn$_3$Sn kagome lattice showing the exact atomic positions (based on Fig. 4a) is overlaid on the d$I$/d$V$ maps. The electronic states taken at –400 mV and –200 mV both form a hexagonal shape, whilst the electronic states at –25 mV form a kagome arrangement, aligning well with the overlaid Mn-based kagome lattice. Thus, confirming the intrinsically localized electronic nature of the flat band.

The combined ARPES/STM study shows definitively that there is a band which is delocalized in momentum space yet non-dispersing to within 150 meV in all three directions, while STM/STS confirms the narrow bandwidth and demonstrates that the band is confined to the Mn kagome lattice and uniform in real space across many lattice constants. The flat band exhibits a bandwidth of 50 meV that can be tuned through controlled Mn defect engineering during growth to just below the Fermi level. The narrow bandwidth, comparable to that of twisted bilayer graphene, indicates truly localized electrons and, consequently, strong electronic correlations. It is found to be robust against varying temperatures, magnetic order, and film thicknesses. Our observation of 3D flat band states induced by strong electronic correlations and quantum confinement, provides a new approach to realizing 3D flat band states in several layer thick kagome magnets. The realization of a robust three-dimensional kagome band in an ultra-thin kagome metal is an important step towards observing kagome-lattice derived, strongly correlated topological and quantum phases when the flat band is tuned to the Fermi level, which in atomically thin kagome systems can be readily achieved via doping, electrostatic-gating, strain effects and/or substrate engineering.

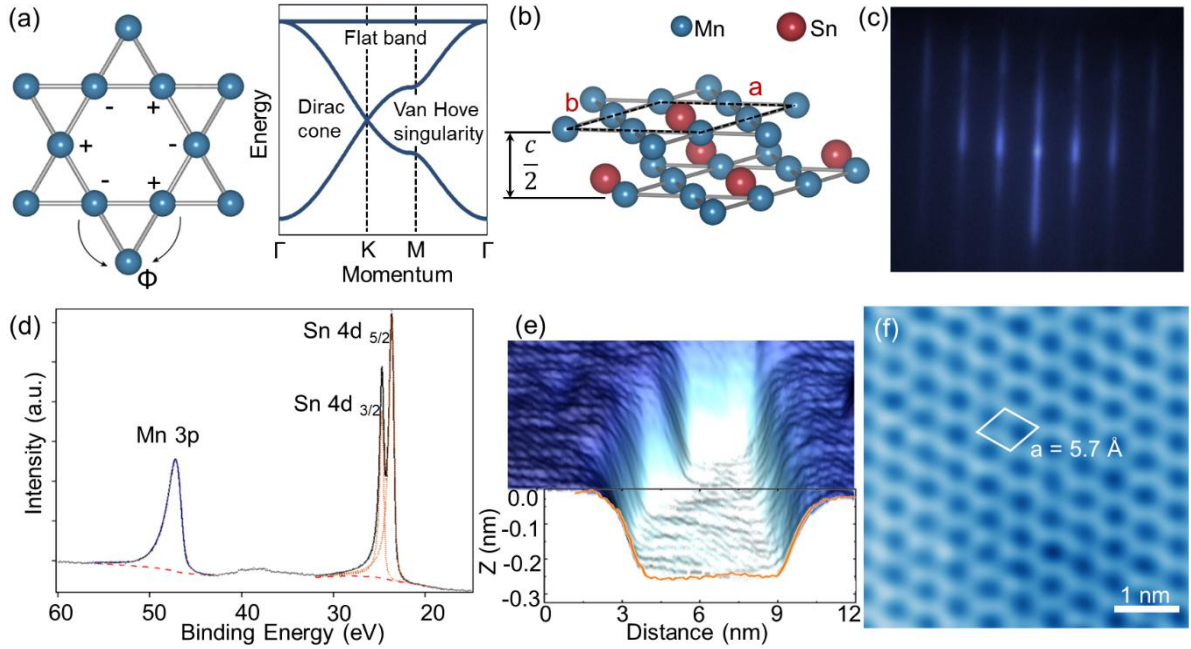

**Figure 1. Structural characterization of ultra-thin Mn₃Sn.** (a) Left panel: Schematic of the kagome lattice with corner-sharing triangles, and flat band resulting from destructive interference (Φ) caused by the cancellation between alternating quantum phases (+, -). Right panel: the tight-binding electronic band structure for the 2D kagome lattice, showing the flat band over the whole Brillouin zone (BZ), Dirac cone at the K point and van Hove singularity at the M point. (b) Lattice structure of Mn₃Sn showing the kagome structure in adjacent layers, where lattice constant $a = b = 0.566$ nm and $c = 0.453$ nm. (c) Reflection high-energy electron diffraction (RHEED) pattern of the epitaxial Mn₃Sn film. (d) X-ray photoemission spectroscopy (XPS) taken at $h\nu = 880$ eV on the Mn₃Sn ultra-thin film, capturing the Mn 3$p$ and Sn 4$d$ core levels. Mn 3$p$ and Sn 4$d$ core levels were fitted by highly asymmetric Voigt functions, presented in blue and orange, respectively. (e) Cross-section scanning tunnelling microscopy (STM) image showing the step edge of adjacent layers of Mn₃Sn film with height of 0.24 nm (bias voltage V = –0.8 V and tunnel current I = 0.1 nA). (f) Atomic resolution STM image taken from the surface of the Mn₃Sn film (bias voltage V = –0.35 V and tunnel current I = 2.2 nA).

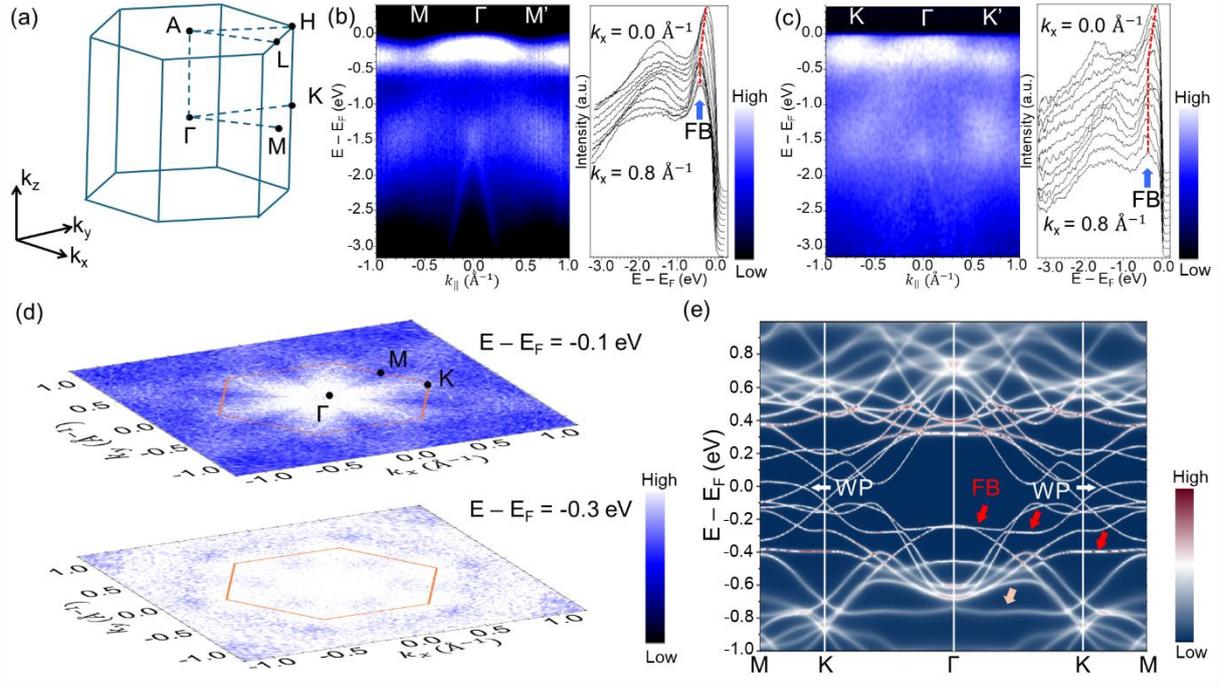

**Figure 2. Observation of kagome flat band in ultra-thin Mn₃Sn.** (a) 3D Brillouin zone of Mn$_3$Sn with marked high-symmetry points for the band structure plots. ARPES spectra and extracted energy distribution curve (EDC) stack plots taken at $T$ = 11 K and $h\nu$ = 104 eV along the (b) ΓM and (c) ΓK directions showing the flat band (label: FB) feature near the Fermi level. The red dashed lines indicate the variation of the flat band peak position as a function of the $k_{\parallel}$ positions, both are smaller than 150 meV. (d) In-plane constant-energy contour maps taken at –0.1 eV and –0.3 eV, measured at $h\nu$ = 104 eV. Red solid line marks the 1$^{st}$ BZ. (e) DFT+DMFT calculated momentum-resolved spectral functions, the colour bar gives the weight of the calculated band. Red/pink and white arrows indicate the flat bands (label: FB) and the Weyl point (label: WP), respectively.

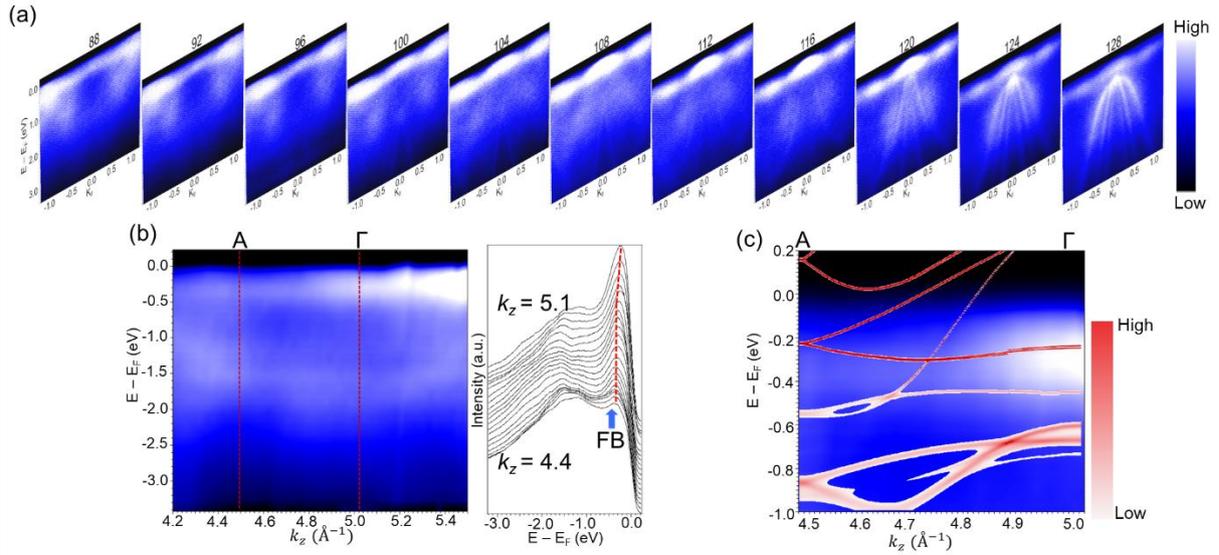

**Figure 3. Confirming the flat band in all three momentum directions.** (a) A series of ARPES spectra taken along the ΓM surface direction at photon energies between 88 eV and 128 eV showing a near 3D flat band feature. (b) Left panel: Out-of-plane ($k_z$) dispersion of the flat band from photon-energy-dependent ARPES experiment with the photon energy tuned from 70 to 120 eV. Right panel: EDC stacks taking from $k_z$ = 4.4 to 5.1 Å$^{-1}$ showing flat band features. The red dashed lines indicate the variation of the flat band peak position as a function of the $k_z$ position, which is smaller than 150 meV. (c) Comparison between ARPES result and DFT+DMFT calculation along with A-Γ direction.

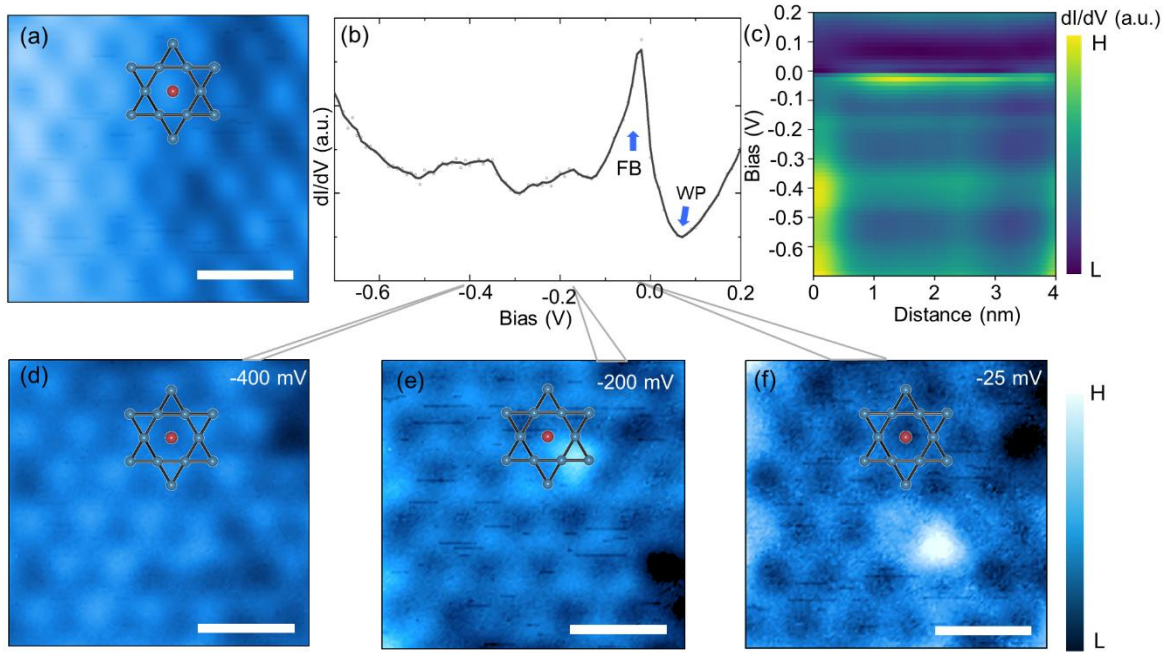

**Figure 4. Imaging the electronic states of the flat band.** (a) Atomic resolution image of a 3x3 nm area of Mn$_3$Sn film surface (image conditions: 0.30 V, 0.5 nA), scale bar is 1 nm. (b) The average scanning tunnelling spectroscopy curve of the whole area in (a). The full width at half maximum (FWHM) of the flat band state is around 50 mV. The arrow points out the Weyl points (label: WP) and flat band (label: FB) in the STS curve. (c) 2D contour plot of a series of average STS curves taken along the horizontal direction in (a). (d)-(h) The STS mappings taken at bias of –400 mV, –200 mV, –25 mV at the same area as (a).

## Methods

### Sample growth and characterization

The epitaxial growth of Mn$_3$Sn ultra-thin film was fabricated using a Scienta Omicron Lab 10 MBE growth chamber with a base pressure of $3\times10^{-10}$ mbar. Before the growth, a Si (111) substrate was flash-annealed at 1180 °C to achieve the 7 × 7 surface reconstruction. The Mn with purity >99% and Sn with purity>99.99% were evaporated from MBE Komponenten effusion cells with deposition rates of 0.045 Å/s and 0.021 Å/s, respectively. The deposition rates were measured by a quartz crystal deposition monitor. During the growth, the silicon substrate was kept between 300 – 320 °C to achieve high-quality growth. Reflection high-energy electron diffraction (RHEED) was used to monitor the growth and determine the initial film quality. The X-ray photoemission spectroscopy (XPS) was conducted on the Pre-vac system at the Soft X-ray spectroscopy beamline station of Australian Synchrotron. The energy resolution was set to 50 meV. The Mn 3*d* and Sn 4*d* core levels were fitted using Voigt profiles. Magnetisation measurements were performed using a Quantum Design MPMS3 SQUID magnetometer. Magnetic fields were applied in the plane of the film. Datasets were normalised to the film volume and corrected for the diamagnetic response of the substrate.

### ARPES measurements

ARPES experiments were performed at beamlines BL03U of the Shanghai Synchrotron Radiation Facility (SSRF)[43,44] and the Soft X-ray spectroscopy beamline of Australian Synchrotron (ANSTO). The BL03U beamline was equipped with DA30 electron analyser with an energy resolution of 20 meV or better. The Soft X-ray spectroscopy beamline was equipped with the Toroidal analyser with an energy resolution of 100 meV. The results were reproduced at both facilities. The angular resolution was set to 0.1° and 0.5° in SSRF and ANSTO, respectively. All samples were transferred from the growth chamber to the beamline station in an ultra-high vacuum suitcase with a base pressure lower than $1\times10^{-9}$ mbar.

### STM & STS measurements

Samples were transferred to Fermi STM on Toroidal analyser and Createc MBE-STM via the ultra-high vacuum suitcase. A Pt-Ir tip was used for the STM and STS measurements. For STS measurements the tip was calibrated on the Au (111) surface showing the Shockley surface state at −0.5 V and flat LDOS near the Fermi level before all measurements. The STS measurements were performed using standard lock-in method with 5-15 mV AC excitation voltage at 823 Hz.

**DFT+DMFT calculations**

We carried out fully charge self-consistent density functional theory plus dynamical mean field theory (DFT+DMFT) calculations using the eDMFT code[45], where the density functional theory part is based on the linearized augmented plane wave method as implemented in WIEN2K package[46]. The Perdew-Burke-Ernzerhof form[47] of the generalized gradient approximation was used for the exchange correlation functional. Hubbard $U = 4.0$ eV and Hund's coupling $J = 0.45$ eV were used in the calculations, the same as in Ref. 27. The DMFT quantum impurity problem was solved by continuous time quantum Monte Carlo method[48,49] at the temperature $T = 116$ K. The formula $U(n_d-1/2)-J(n_d-1)/2$ ($n_d$ is the nominal occupation of the Mn $3d$ electrons) was used to subtract the double counting energy in the DFT+DMFT formula. More details are included in Ref. 27.

**Acknowledgement**

J.B., M.Z., G.L.C, M.S.F., A.T. and M.T.E. acknowledge funding support from ARC Centre for Future Low Energy Electronics Technologies (FLEET) CE170100039. M.T.E. acknowledges funding support from ARC Future Fellowship FT220100290. M.Z., A.T., M.T.E. acknowledge funding support from ANSTO-Postdoc Fellowship. M.Z. and G.L.C. acknowledge funding support from AINSE Early Career Research Grant. J.B. and M. B. acknowledge this research was supported by an AINSE Ltd. Postgraduate Research Award (PGRA). Z.Y. is supported by the Fundamental Research Funds for the Central Universities (Grant No. 2243300003), the Innovation Program for Quantum Science and Technology (Grant No. 2021ZD0302800), the National Natural Science Foundation of China (Grants No. 12074041). Part of this research was undertaken on the soft X-ray beamline at the Australian Synchrotron, part of ANSTO. The high-resolution ARPES measurements with micro-focused beam were performed at Beamline 03U of Shanghai Synchrotron Radiation Facility, which is supported by SiP.ME$^2$ project under Contract No. 11227902 from National Natural Science Foundation of China (NSFC). The calculations were carried out on the high-performance computing cluster of Beijing Normal University in Zhuhai.

**Author contributions**

M. T. E, and M. Z. devised the experiments. J. B and M.Z performed the MBE growth and STM/STS measurements at Monash. M.Z. J. B, A. T. and M.T.H. B performed the ARPES and XPS measurements at the Australian Synchrotron. M.Z. J. B, H. Z. performed the ARPES measurements at the Shanghai Light Source with the support from J. L, W. J, Z.-T. L. and M. Y. The magnetic measurements were performed by G. L. C. The DFT calculations were performed by T. Y and Z. Y. M. T. E., M. Z. and J. B composed the manuscript. All authors read and contributed feedback to the manuscript.